\documentclass[twocolumn,twoside,slac]{revtex4}
\usepackage{graphicx}
\usepackage{fancyhdr}
\usepackage{amssymb,amsmath}    

\newcommand{\GeantF} {{\tt Geant4}}
\newcommand{\GeantT} {{\tt Geant3}}
\newcommand{\GtoG}   {{\tt G3toG4}}

\newcommand{\FLUKA}  {{\tt FLUKA}}

\newcommand{\ROOT}   {{\tt ROOT}}

\newcommand{\AliRoot}{{\tt AliRoot}}

\newcommand{\GHEISHA}{GHEISHA}
\newcommand{\Energy}  {\ensuremath{\mathrm{E}}} 
 
\newcommand{\Mass}    {\ensuremath{\mathrm{M}}}            
\newcommand{\cm}     {\ensuremath{ \,{\mathrm{cm}}}}

\newcommand{\degree} {\ensuremath{ ^\circ}}

\newcommand{\MeV}    {\ensuremath{ \,{\mathrm{MeV}}}}
\newcommand{\GeV}    {\ensuremath{ \,{\mathrm{GeV}}}}

\newcommand{\Let}    {\ensuremath{ \,{\mathrm{L}}}}
\newcommand{\microC} {\ensuremath{ \,{\mu \mathrm{C}}}}
\newcommand{\n}      {\ensuremath{ \,{\mathrm{n}}}} 

\newcommand{\photon}  {\ensuremath{ {\gamma}}}

\newcommand{\elp}     {\ensuremath{ {\mathrm e}^+}}

\newcommand{\elpm}    {\ensuremath{ {\mathrm e}^\pm}}

\newcommand{\sumatorio}[2]  {\underset{#1}{\overset{#2}{\sum}}}

\newcommand{\NPA}[4]{#1, Nucl. Phys. A #2 (#3) #4}
\newcommand{\NSE}[4]{#1, Nucl. Sci. Eng. #2 (#3) #4}

\begin{document}

\title{ALICE experience with GEANT4}

%

\author{I. Gonz\'alez Caballero}
\affiliation{Instituto de F\'{\i}sica de Cantabria, Santander, Spain}
\author{F. Carminati}
\affiliation{CERN, Geneva, Switzerland}
\author{I. H\v{r}ivn\'{a}\v{c}ov\'{a}}
\affiliation{IPN, Orsay, France}
\author{A. Morsch}
\affiliation{CERN, Geneva, Switzerland}

\begin{abstract}
Since its release in 1999, the LHC experiments have been evaluating
\GeantF\ in view of adopting it as a replacement for the obsolescent
\GeantT\ transport MonteCarlo. The ALICE collaboration has decided to
perform a detailed physics validation of elementary hadronic processes
against experimental data already used in international benchmarks. In
one test, proton interactions on different nuclear targets have been
simulated, and the distribution of outgoing particles has been
compared to data. In a second test, penetration of quasi-monoenergetic
low energy neutrons through a thick shielding has been simulated and
again compared to experimental data. In parallel, an effort has been
put on the integration of \GeantF\ in the \AliRoot\ framework. An
overview of the present status of ALICE \GeantF\ simulation and the
remaining problems will be presented. This document will describe in
detail the results of these tests, together with the improvements that
the \GeantF\ team has made to the program as a result of the feedback
received from the ALICE collaboration. We will also describe the
remaining problems that have been communicated to \GeantF\ but not yet
addressed.

\end{abstract}

\maketitle

\thispagestyle{fancy}


\section{ALICE detector}

ALICE ~\cite{ALICE}, {\it A Large Ion Collider Experiment}, is one of the four
experiments that will run in the LHC ({\it Large Hadron
Collider}). Specially designed for heavy-ion physics its main aim is
to study the properties of strongly interacting matter at extreme
energy densities, where the formation of the quark-gluon plasma is
expected.

\section{Geant4}

\GeantF~\cite{GEANT4} is the successor of the very popular and
successful simulation package \GeantT~\cite{GEANT3}. \GeantF\ has been
completely written in C++ using Object Oriented technologies to
achieve a greater degree of transparency, flexibility and
extensibility. Its target application areas include high energy
physics, as well as nuclear experiments, medical, accelerator and
space physics studies. Thanks to its rich set of physics processes it
is expected to supersede the \GeantT\ package, which is not updated
anymore.

\section{ALICE Software}

\subsection{AliRoot}
\begin{figure}[htb!]
  \centering
  \includegraphics[width=\linewidth]{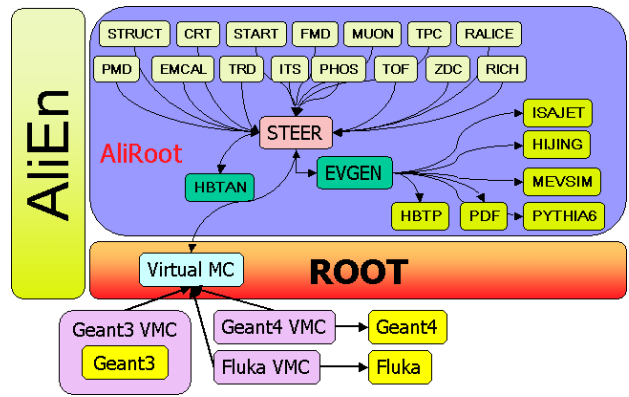}
  \caption{\sl Schematic view of the \AliRoot\ internal structure.}
  \label{fig:AliRoot}
\end{figure}

\AliRoot\ is the ALICE off-line framework for simulation,
reconstruction and analysis. It uses the \ROOT~\cite{ROOT} system as a
foundation on which the framework and all applications are built. For
the simulation both the actual transport code (see subsection
\ref{sec:VMC}) and the generators have been factorized into abstract
classes so that changing between all the possible choices can be done
very easily. A schematic view of the internal structure of \AliRoot\
is shown in figure \ref{fig:AliRoot}.

\subsection{Virtual Monte Carlo} \label{sec:VMC}

The Virtual Monte Carlo (VMC), thoroughly described in another paper
at this conference~\cite{VMC}, is a simulation framework developed
within the ALICE collaboration in close contact with the \ROOT\
team. It is based on the \ROOT\ system and isolates the code needed to
perform a given detector simulation from the real transport code that
is used. Once the user application is built, changing between any of
the MCs can be done by changing just one line of code in a \ROOT\
macro. \GeantT\ and \GeantF\ are currently fully integrated in the
VMC. The integration of \FLUKA\ is almost finished.

\begin{figure}[htb!]
  \centering
  \includegraphics[width=\linewidth]{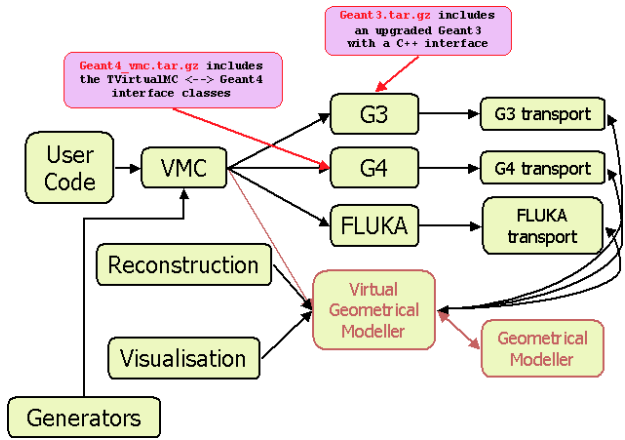}
  \caption{\sl Schematic view of the experimental software framework 
               based on the VMC and the Geometrical Modeller.}
  \label{fig:VMC}
\end{figure}

The VMC is now being integrated with the new, more neutral and
efficient, Geometrical modeller developed for HEP by the ALICE Offline
project in a close collaboration with the ROOT team. The native
geometry package in each MC is going to be replaced with this new
geometry package.  Figure \ref{fig:VMC} shows the structure of the
experiment software framework based on the VMC and the Geometrical
Modeller.

The \GeantF\ implementation of the VMC, the Geant4 VMC, is completed
and full simulation is posible even in complex geometries. Among the
characteristics in Geant4 VMC the following can be found:

\begin{itemize}

\item It is a layer between VMC and \GeantF\, independent of the ALICE
software.

\item It provides an easy and tranparent transition from a \GeantT\ based
simulation to a \GeantF\ based simulation.

\item It uses the \GtoG\ package (included in \GeantF\ and developed with
a substantial contribution by ALICE), so it provides full support for
reflections and limited support for \GeantT\ geometries using ``MANY''
option.

\item It has the capability to switch between the \ROOT\ user
interface and the \GeantF\ user interface, processing foreign commands
or macros in both UIs.

\item It includes a \GeantF\ geometry browser.

\item It has an XML export module.

\end{itemize}

The VMC examples, provided with the VMC~\cite{VMCweb}, allow
comparisons between all the MC implementations. Besides, \AliRoot\ is
an example of the possible use of the VMC concept in a complex HEP
application.

\subsection{ALICE Geant4 Simulation}

At present, the ALICE simulation code based on the VMC can be run with
\GeantF\ including the geometry of all 12 subdetectors and all
structure modules. However, two detector subsystems are still excluded
from hits production: the first one (ITS) since it uses not yet
supported ``MANY'' positions, and the second one (RICH) because of a
special requirement for adding its own particles to the stack (not yet
available in the Geant4 VMC).

Runs of 5000 primary particles with the HIJING parameterisation event
generator (representing 5.8~\% of a full background event) and with
the standard AliRoot magnetic field map were performed. The kinetic
energy cuts equivalent to those in \GeantT\ simulations were applied,
using a special process and user limits objects. The hit distributions
in x and z distributions were compared for all included subdetectors
and for \GeantT\ and \GeantF. They were found to be compatible.

%
%
%
%
\section{Hadronic Benchmarks}
\subsection{Reasons}
In the context of ALICE, \GeantF\ is considered as a possible
replacement for \GeantT. The hadronic processes are of capital
importance for the physics that will be studied in this detector.

Low momentum particles are of great concern for the ALICE detectors
covering the central rapidity zone and the forward muon spectrometer
since ALICE has a rather open geometry with no calorimetry to absorb
particles and a small magnetic field. At the same time low momentum
particles appear at the end of hadronic showers. Residual background
which limits the performance in central Pb-Pb collisions results from
particles "leaking" through the front absorbers and beam-shield.

Therefore, we are performing a set of benchmarks of the hadronic
\GeantF\ processes against experimental data.

\subsection{Proton thin-target benchmark}
 The proton thin-target benchmark aims at establishing the
capabilities of \GeantF\ to reproduce the single hadronic interactions
on nuclei in the so called intermediate energy range
($100~\MeV~<~\Energy_{lab}~<~1~\GeV$).

These studies were started with the release 3.0 of the \GeantF\
code. During the running of the benchmark we experienced several
problems including some crashes of the program. They were reported to
the \GeantF\ team.  In \cite{ALICE-note} we published the results
obtained with the official release 3.2 . Our current revision of these
results was obtained with the latest release available at the time of
the conference: 5.0 with patch 1.

\subsubsection{Experimental setup}
\begin{figure}[hbt!]
\centering
  \includegraphics[width=\linewidth]{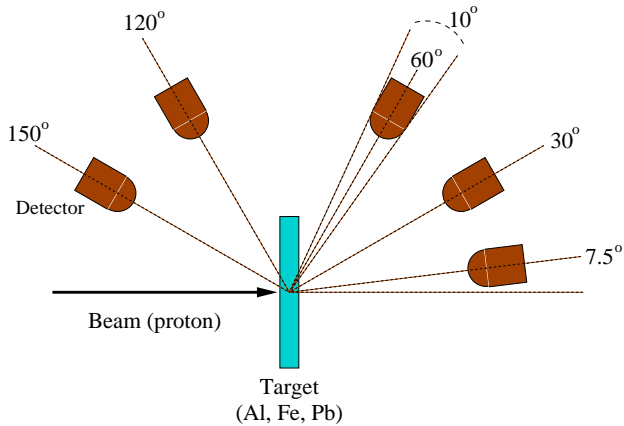}
  \caption{\sl Schematic description of the experimental setup.}
  \label{fig:experiment}
\end{figure}

The data used in this benchmark was collected at Los Alamos National
Laboratory (New Mexico, USA) in the Weapons Neutron Research Facility.
A schematic representation of the experimental setup is shown in
Figure \ref{fig:experiment}. A proton beam is directed towards a thin
target, therefore no more than one collision is expected for most of
the cases. The selected target materials are aluminum, iron and lead.
The proton energies vary from 113 \MeV\ to 800 \MeV.  Neutrons are
detected at several polar angles ranging from 7.5\degree\ to
150\degree.  A detailed description of the experiment can be found
in~\cite{experiment}.

\subsubsection{\GeantF\  simulation}

Due to the low hadronic interaction cross section, the probability of
having a single interaction in a thin target is small. Most of the
times, the protons traverse the target material without interacting.
For this reason, to speed up computation, a setup different from the
real one was simulated with \GeantF. A big box made of the target
material was built to make sure that one hadronic interaction would
take place.  Only transportation and proton inelastic processes (class
{\tt G4ProtonInelasticProcess}) were activated. Immediately after the
interaction, the kinematic properties of the secondaries produced were
stored for further analysis, and the next primary interaction was
generated. The direction of each neutron produced was compared with
the position of the detectors in the experimental setup.

Two physics models inside \GeantF\ were used for this benchmark:
\begin{enumerate}
\item The first model studied is the Parameterized model. This is more
  or less the \GeantF\ implementation of the \GHEISHA~\cite{GHEISHA} model
  (\mbox{\tt G4LEProtonInelastic} and {\tt G4HEProtonInelastic}
  classes). Some improvements with respect to the \GeantT\
  implementation of the model are claimed. As it is a parameterized
  model, the nuclear fragments remaining from the inelastic collision
  are not calculated. To be able to verify the model we have deduced
  the fragment properties from the known conservation laws.
  
\item The second model used is the \GeantF\ implementation of the
  precompound~\cite{Precompound} model ({\tt G4PreCompoundModell}
  class). It is a microscopic model which is supposed to complement
  the hadron kinetic model in the intermediate energy region for
  \mbox{nucleon--nucleus} inelastic collisions. From \GeantF\ release
  5.0 a new intranuclear cascade model is available in \GeantF\ ({\tt
  G4CascadeInterface} class). The latest studies were done with this
  model activated in our code.

\end{enumerate}

Finally, 200K events were generated for each model, energy and
material, and the full statistics was used in the following studies.


\subsubsection{Consistency checks}
The first exercise that we did consisted in a set of consistency
checks, namely conservation laws and azimuthal distributions.

\begin{figure}[!h]
  \centering
  \includegraphics[width=\linewidth]{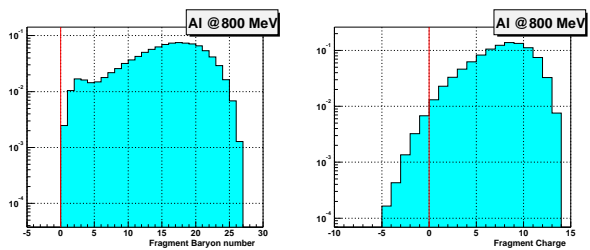}
  \caption{\sl Charge and baryon number distributions (normalized to 1) in the residual fragment from conservation laws, for protons at 800 \MeV\ on aluminum (Parameterized model).}
  \label{fig:AlBaryonBalance}
\end{figure}

There are four systems in the reaction: the incident proton, the
target nucleus, the emitted particles and the residual fragments. Four
fundamental conservation laws can be checked: Energy, momentum, charge
and baryon number. The parameterized model does not generate a
residual fragment making these checks impossible. However the
fundamental correlation laws allow us to determine the energy and the
momentum of the of residual fragments, and hence the square of its
total mass, while barion and charge conservation can give us the
number of protons and neutrons in the fragments. Performing the
calculation, we found that up to 1.5\% of the events have unphysical
states with a residual having a negative number of protons or neutrons
(see table \ref{tab:NegativeBaryon} or figure
\ref{fig:AlBaryonBalance}), or with a negative value of $\Mass^2$.

\begin{table}[htb!]
  \begin{center}
    \begin{tabular}{|c|c||c|c|c|}
      \hline
      Material  & Energy   & $Q < 0$ & $B < 0$ & $N_{neu} < 0$ \\
      \hline
      \hline
                 & 113 \MeV & 0.00 \% & 0.00 \% & 0.00 \%\\
       Aluminum & 256 \MeV & 0.33 \% & 0.02 \% & 0.44 \% \\
                 & 597 \MeV & 0.76 \% & 0.00 \% & 0.90 \% \\
                 & 800 \MeV & 1.20 \% & 0.00 \% & 1.50 \% \\
       \hline
                 & 113 \MeV & 0.00 \% & 0.00 \% & 0.00 \% \\
                 & 256 \MeV & 0.00 \% & 0.00 \% & 0.00 \% \\
       Iron      & 597 \MeV & 0.01 \% & 0.00 \% & 0.02 \% \\
                 & 800 \MeV & 0.01 \% & 0.00 \% & 0.05 \% \\
       \hline
                 & 113 \MeV & 0.00 \% & 0.00 \% & 0.00 \% \\
                 & 256 \MeV & 0.00 \% & 0.00 \% & 0.00 \% \\
       Lead      & 597 \MeV & 0.00 \% & 0.00 \% & 0.00 \% \\
                 & 800 \MeV & 0.00 \% & 0.00 \% & 0.00 \% \\
      \hline
    \end{tabular}
  \end{center}
  \caption{{\sl Percentage of events having a fragment with unphysical values of charge (number of protons), baryon number and number of neutrons (Parameterized model).}}
  \label{tab:NegativeBaryon}
\end{table}

\begin{figure}[!h]
  \centering
  \includegraphics[width=\linewidth]{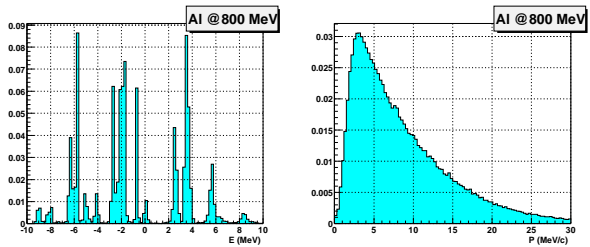}
  \caption{\sl Energy and momentum balance (i.e. non-conservation) for protons at 800 \MeV\ on aluminum (Precompound + Cascade models).}
  \label{fig:AlEPPrec}
\end{figure}

Some violations of the conservation laws were observed in the
initially tested versions of \GeantF, where the precompound model had
to be used alone (there was no cascade model). They were solved and
only a small energy non-conservation remained, apparently coming from
the final de-excitation phase of the model. Surprisingly when adding
the cascade model available since the 5.0 release we observed that
neither momentum, nor energy are conserved as can be seen in figure
\ref{fig:AlEPPrec}.

Azimuthal distributions had a known bug in the old \GeantT\
implementation of \GHEISHA. For this reason we checked them for the
parameterized and precompound model finding several azymuthal
asymmetries in the first versions of \GeantF. The latest release
corrects all of them and the distributions are flat as expected.

\subsubsection{Double differentials}
\begin{figure}[!h]
  \centering
  \includegraphics[width=\linewidth]{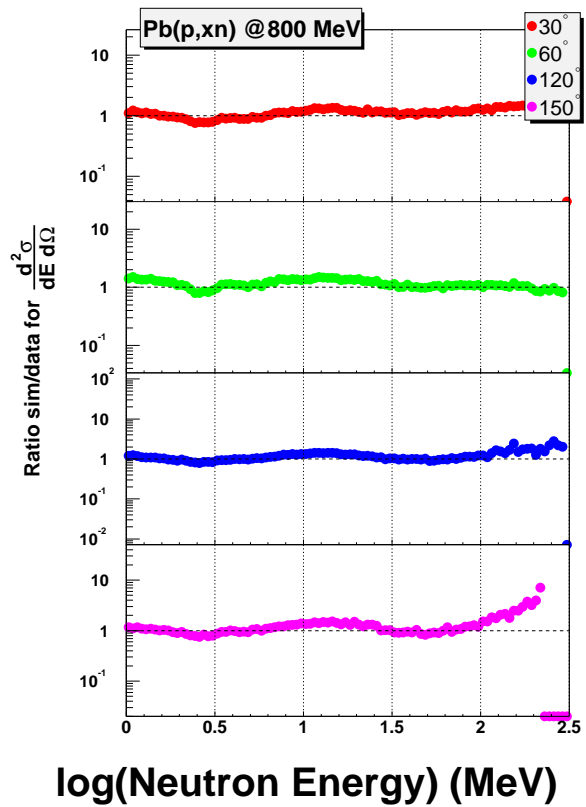}
  \caption{\sl $\frac{d^2 \sigma}{d\Energy d\Omega}$ ratio between MC and data for protons at 800 \MeV\ on lead (Precompound + Cascade models).}
  \label{fig:DDRatioPb800Pre}
\end{figure}

\begin{figure}[!h]
  \centering
  \includegraphics[width=\linewidth]{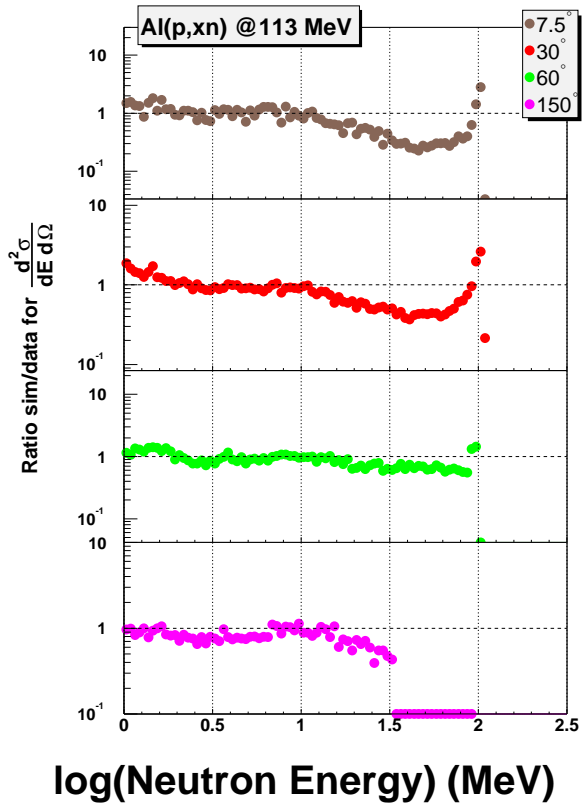}
  \caption{\sl $\frac{d^2 \sigma}{d\Energy d\Omega}$ ratio between MC and data for protons at 800 \MeV\ on aluminum (Precompound + Cascade models).}
  \label{fig:DDRatioAl113Pre}
\end{figure}

The double differential distribution, $\frac{d^2 \sigma}{d\Energy d
  \Omega}$, of the emitted neutrons was calculated for all the
  cases. The parameterized model is not able to correctly reproduce
  most of the distributions. The same applies to the precompound model
  alone. Adding the cascade model improves a lot the agreement between
  data and MC (see figure \ref{fig:DDRatioPb800Pre}), though some
  discrepancy still persists for low incident energy and light
  targets (see figure \ref{fig:DDRatioAl113Pre}).


\subsection{Neutron transmission benchmark}
Here we describe a second benchmark on the neutron transport inside
iron and concrete in the very low energy region
($10~\MeV~<~\Energy_{lab}~<~100~\MeV$). For this benchmark the 4.1
(patch 1) release of \GeantF\ was used.

\subsubsection{Experimental setup}
In this test we use the data coming from an international
benchmark~\cite{TIARA} that took place at the TIARA facility of JAERI.

\begin{figure}[hbt]
    \centering
    \includegraphics[width=\linewidth]{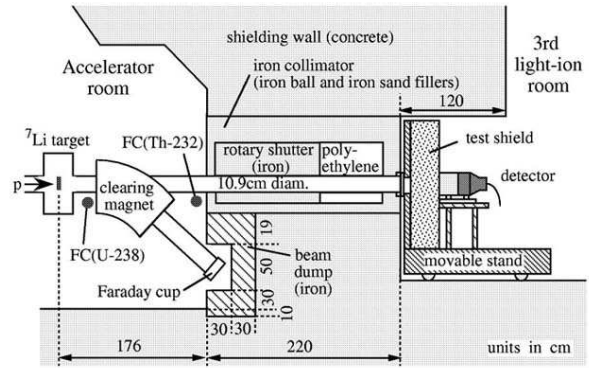}
    \caption{\sl Experimental Setup for the neutron transmission experiment at the TIARA facility.} 
    \label{fig:ExperimentalSetup}
  \end{figure}

A cross section view of the TIARA facility with the experimental
arrangement can be seen in figure
\ref{fig:ExperimentalSetup}. 43~\MeV\ and 68~\MeV\ protons bombard a
$^7$Li target, producing a \mbox{quasi-monoenergetic} source of
40~\MeV\ and 65~\MeV\ neutrons. Iron and concrete shields of different
widths were placed 401~\cm\ away from the neutron source. The neutron
flux was measured in three different positions on the horizontal
direction: at 0, 20 and 40~\cm\ from the beam axis.

The details on how the neutron intensity and energy spectra was
measured can be found in \cite{TIARA}.

\subsubsection{\GeantF\ simulation}
There are two main parts which define the \GeantF\ simulation: the
definition of the geometry and the choice of the appropriate set of
active physic processes, i.e. the so called {\em physics list}.

\begin{figure}[hbt]
   \centering 

   \includegraphics[height=\linewidth,angle=270]{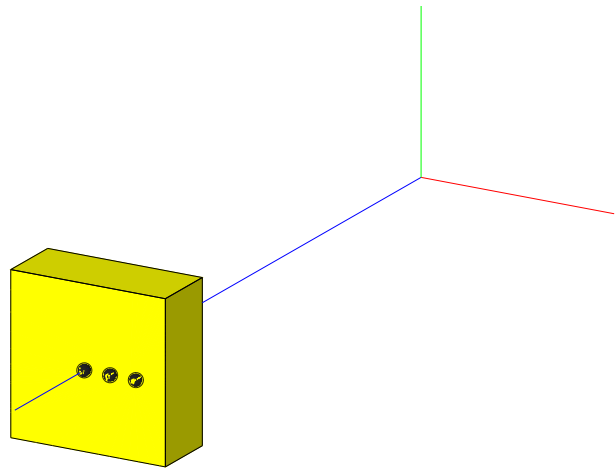}
   \caption{{\sl Simulation setup as obtained directly from \GeantF.}}
   \label{fig:SimulationSetup} 
\end{figure}

The full geometry of the experimental setup, as described in the
previous section was not implemented for the simulation. A simpler
instance (see figure \ref{fig:SimulationSetup}) was used. A block of
the material being studied was built at 401~\cm\ from the source in
the z direction, inside a big vacuum hall. In order to measure the
flux three empty spherical sensitive detectors were constructed just
behind the target at 0~\cm, 20~\cm\ and 40~\cm\ from the beam axis in
the x direction.

The energy and angular neutron spectra of the source was
simulated. For the energy, the distribution obtained from the measures
after the lithium target was used. The input spectrum was reproduced
accurately.

A selection of the available processes was done in order to match our
physics requirements. The physics list was divided in five major
blocks:
\begin{enumerate}
  \item{General processes}: This block includes only the decay
  process. It was activated only for neutrons.

  \item{Electromagnetic processes}: The electromagnetic processes were
  activated only for \photon, \elpm, protons and alpha particles. The
  processes activated were, for all mentioned particles, multiple
  scattering and ionization. For photons, the photo electric effect,
  the Compton scattering, and the \photon\ conversion were
  added. Bremsstrahlung was included for \elpm\ and \elp\ annihilation
  was included for \elp.

  \item{Hadronic elastic processes}: These processes were switched on
  for protons, neutrons, deuterons, tritons and alphas only. The
  default low energy hadronic elastic model (class {\tt G4LEElastic})
  was used alone, except for neutrons where its limit of applicability
  was set to 19~\MeV\ and the more specialized high precision model
  (class {\tt G4NeutronHPElastic}) was also activated.

  \item{Hadronic inelastic processes}: The same particles as in the
  previous block had these kind of processes activated. The
  precompound model was selected. For neutrons, the cross-sections
  provided by the \GeantF\ class {\tt G4NeutronInelasticCrossSection}
  were used together with the inelastic high precision model below
  19~\MeV.

  \item{Other hadronic processes}: These are only special neutron
  processes like neutron fission and neutron capture. The default
  models, together with their high precision version for energies
  below 19~\MeV, were used.
\end{enumerate}

\subsubsection{Flux Estimation}
The track length method was used to estimate the flux after the
target. Three spheres of 5.08~\cm\ filled with vacuum were placed
tangent to the target at x~=~0~\cm, x~=~20~\cm\ and x~=~40~\cm, and
y~=0, being z the axis of the beam perpendicular to the shielding. For
every neutron entering each sphere its entry point, exit point,
energy, $\Energy$, and track length, $\ell(\Energy)$, inside the
volume of the sphere ($V_S$) is stored. The flux in a given energy
interval, $\Delta \phi (\Delta \Energy)$, is then calculated as the
track length normalized with the sphere volume. So for $N$ events we
have:

\begin{equation*}
\Delta \phi (\Delta \Energy) = \frac{\sumatorio{\Energy \epsilon \Delta \Energy}{}{\ell(\Energy)}}{V_s \cdot N}
\end{equation*}

The final quantity takes into account the intensity, $I$, of the
incident flux, and is normalized with the lethargy, $L = \Delta \log
(\Energy)$. Therefore, the final quantity studied becomes:

\begin{equation*}
\phi (\Energy) [\n \cdot \cm^{-2}\cdot \Let^{-1} \cdot \microC^{-1}] = 
	\frac{I \cdot \sum{\ell(\Energy)}}{V_s \cdot N \cdot \Delta \log (\Energy)}
\end{equation*}

\subsubsection{Results}

\begin{figure}[hbt]
  \centering
  \includegraphics[width=\linewidth]{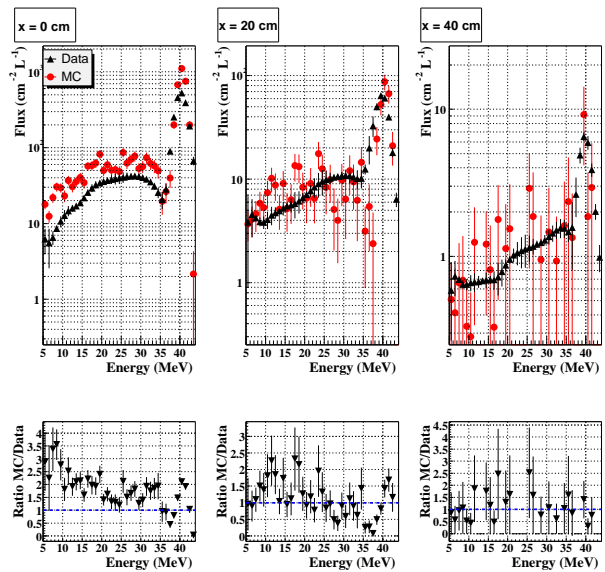}
  \caption{\sl Flux distributions for neutrons at 43 \MeV\ traversing a 40~\cm\ iron block.} 
  \label{fig:Flux1}
\end{figure}

\begin{figure}[hbt]
  \centering
  \includegraphics[width=\linewidth]{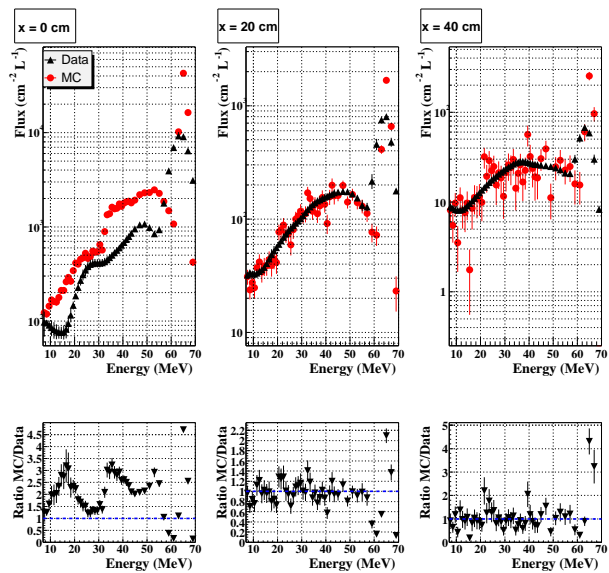}
  \caption{\sl Flux distributions for neutrons at 68 \MeV\ traversing a 50~\cm\ concrete block.} 
  \label{fig:Flux2}
\end{figure}

We remark a large and consistent discrepancy between experimental data
and \GeantF\ results. In particular we notice how the transmission
peak is overestimated by \GeantF\ and there seems to be an
overestimation of the fluency again around 15 \MeV. These differences
might come from the simplified geometry and further investigation is
needed. See figures \ref{fig:Flux1} and \ref{fig:Flux2} for some
results.


All these results have been communicated to the \GeantF\ experts as
they were produced.

\subsection{Conclusions}
Two sets of experimental data on proton-thin target and neutron
transmission experiments were used to benchmark \GeantF. During the
exercise several inconsistencies were found and reported to the
\GeantF\ experts. This fact, together with the lack of some models in
the MC toolkit limited the precision with which the experimental
results could be reproduced. Most of the problems are now corrected
though it seems there is still space for more improvement, specially
in the energy-momentum conservation laws of the recently added cascade
model.

\section{G4UIRoot}
G4UIRoot~\cite{G4UIRoot} is a new \GeantF\ GUI built using the \ROOT\
libraries. It fully integrates into \GeantF\ and brings together the
strengths of the \GeantF\ user interface and \ROOT\ capabilities.

\subsection{Motivations}
\GeantF\ has already several GUIs (Xm, OPACS, GAG). Nevertheless, none
of them was found completely satisfactory. Xm is more a UI, and the
fact that uses Motif gave rise to several problems on different
platforms. The OPACS GUI was difficult to integrate in a program, and,
though it seems extremely customizable, working with it was not found
comfortable. Java GAG was very user-friendly. However the fact that it
was written in Java made it slow.

On the other hand the growing \ROOT\ community feel comfortable with it
and gets used to its particular way of interacting with the programs
through the C++ interpreter. This GUI aims at bridging both worlds
allowing at the same time access to the \GeantF\ sophisticated high
level commands and to the code itself through the \ROOT\ interpreter.

\begin{figure*}[!hbt]
  \centering
  \includegraphics[width=0.8\linewidth]{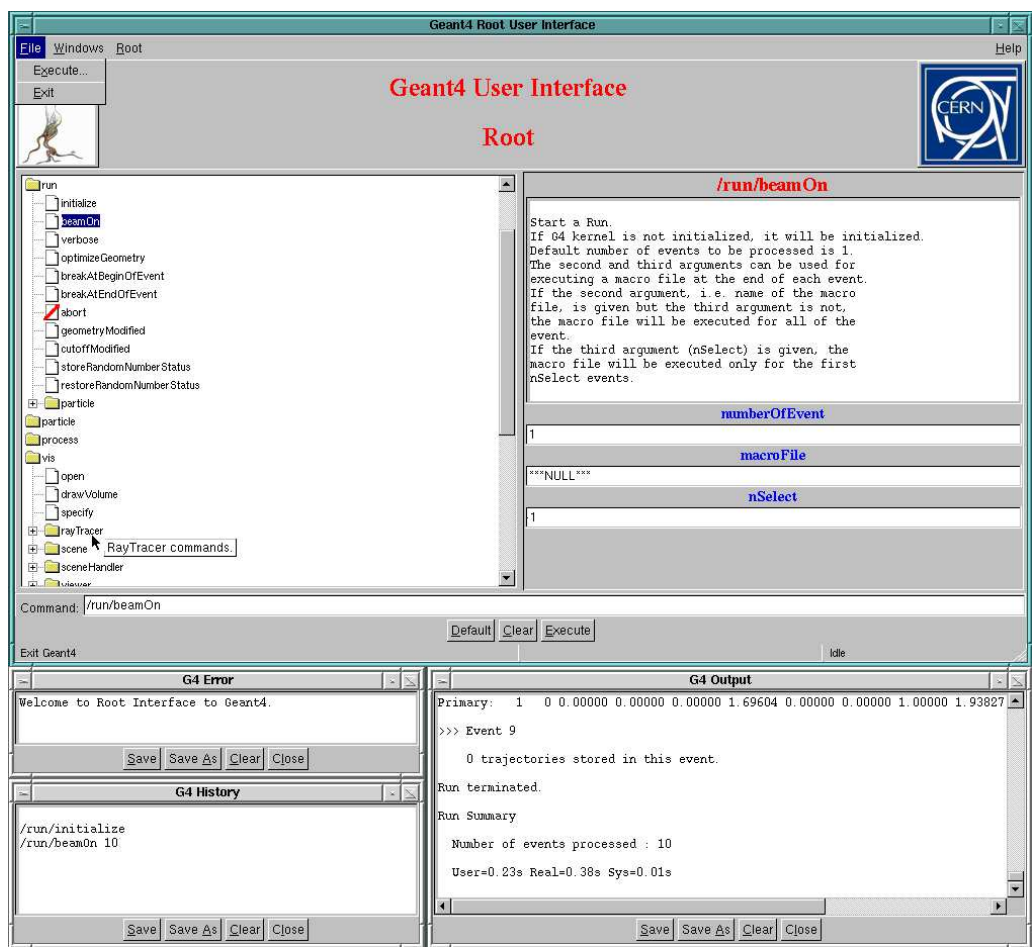}
  \caption{{\sl G4UIRoot main window at the top. The normal output window is at the bottom right. On the bottom left the history window is shown and above it the error window.}}
  \label{fig:G4UIRoot}
\end{figure*}

A capture of the main G4UIRoot window together with some output
windows can be seen in figure \ref{fig:G4UIRoot}. It is easy to
realize that this GUI is highly influenced by GAG. Its look and feel
is similar, and, although many enhancements have been added, the basic
functionality is common. GAG code was taken as a starting point and
showed very helpful.

\subsection{Features}
This GUI was constructed using the \ROOT\ toolkit and profiting from
its GUI capabilities. It is fully integrated into the \GeantF\
compilation framework. Some of its features are:
\begin{itemize}

\item Command tree: The full command tree can be inspected in a
tree-like structure where the availability of the commands is clearly
marked and updated according to the \GeantF\ state.

\item Direct command typing: Commands may be executed by directly
typing them in a space at the bottom of the main window. This is very
much like the normal \GeantF\ user interface with some extensions
(tabbing, history). Typing a command here updates the selection in the
command tree.

\item Parameters frame: The list of parameters for a given selected
command is displayed in a frame with the default values and the
possibility of modifying them.

\item Command help: The full command help is also displayed in the
parameters frame and the short command help appears in pop-up windows
and the status bar.

\item Main window customization: Pictures and titles in the main
window may be customized. \GeantF\ also provides a way to add new
menus to access already registered \GeantF\ commands.

\item Access to external \GeantF\ macros and \ROOT\ {\tt TBrowser}.

\item Output windows: Normal and error output are shown in different
windows with saving capabilities.

\item History: History is logged to another window and may be
saved. It may also be recalled at any point from the command line.

\item \ROOT\ interpreter: The terminal from which the application is
launched runs the typical \ROOT\ interpreter. It provides run-time
access to the objects for which the \ROOT\ dictionaries were generated
(all \ROOT\ objects and the user objects based on the \ROOT\
framework). For the time being, this is not the case for \GeantF\
objects.

\end{itemize}

\subsection{Conclusions}
The first version of G4UIRoot was developed in a very short time
thanks to both the good desing of the \ROOT\ GUI and the \GeantF\
interface categories. A useful GUI for new-comers, people used to
\ROOT\ and interactive users is now available. It has most of the
capabilities of other \GeantF\ user interfaces and some more
extensions not found in any of them. Some interest has been shown from
other people and very usefull contributions have been provided from
some of them.

\end{document}